\documentclass[pre,twocolumn,nofootinbib,amsmath,amssymb,]{revtex4-2}
\usepackage{eucal}
\usepackage{graphicx}   
\usepackage{verbatim}   
\usepackage{color}      
\usepackage{subfigure}  
\usepackage{hyperref}   
\usepackage{soul}
\usepackage{hyperref}
\usepackage{tabularx}
\usepackage{amsmath,blkarray}
\usepackage{float}
\usepackage{amssymb}
\usepackage{mathtools}
\usepackage{setspace}
\usepackage{amsmath}
\usepackage{amsfonts}
\usepackage{bm}
\usepackage{colortbl}
\usepackage{outlines}
\usepackage{xcolor,cancel}
\usepackage{subfigure}
\usepackage[utf8]{inputenc}
\usepackage[T1]{fontenc}
\usepackage{mathptmx}
\usepackage{dcolumn}
\usepackage{float}
\usepackage{tcolorbox}

\usepackage{graphicx}
\usepackage{multirow}

\begin{document}
	\author{Soumen Das}
	\email{soumen.das@tifr.res.in }
	\affiliation{Department of Condensed Matter Physics and Materials Science,\\ Tata Institute of Fundamental Research, Mumbai 400005, India}
		\author{Shankar Ghosh}%
	\email{sghosh@tifr.res.in}
	\affiliation{Department of Condensed Matter Physics and Materials Science,\\ Tata Institute of Fundamental Research, Mumbai 400005, India}%
	\author{Shamik Gupta}
	\email{shamikg1@gmail.com}
	\affiliation{Department of Physics, Ramakrishna Mission Vivekananda
		Educational and Research Institute, Belur Math, Howrah 711202, India\\ Regular Associate, Quantitative Life Sciences Section, ICTP - The Abdus Salam International Centre for Theoretical Physics, Strada Costiera 11, 34151 Trieste, Italy}
	\date{\today}
	\title{Self-similar inhomogeneous stationary states under constrained dynamics}
	\date{\today}
	\begin{abstract}
	 The dynamics of $n$ rigid objects, each having $d$ degrees of freedom, is played out in the configuration space  of dimension $nd$. Being rigid, there are additional constraints at work that render a portion of the configuration space inaccessible. In this paper, we make the assertion that treating the overall dynamics as a Markov process whose states are defined by the number of contacts made between the rigid objects provides an effective coarse grained characterization of the otherwise complex phenomenon. This coarse graining reduces the dimensionality of the space from $nd$ to one. We test this assertion for a one dimensional array of curved squares each of which
    is undergoing a biased diffusion in its angular orientation.
	 \end{abstract}
	\maketitle
	\section{Introduction}
It is widely recognized that quenched spatial disorder has rather nontrivial consequences on the long-time behavior of a system of interacting constituents driven by an external field, resulting in such intriguing and complex phenomena as macroscopic density inhomogeneity leading to phase separation~\cite{barma2006} and a non-monotonic field-induced drift velocity as a function of the field strength~\cite{white1984field}. While unveiling of such fascinating effects has mostly been for the case in which the interacting constituents are point particles, the issue of whether similar effects may be observed with a collection of extended objects being driven by a field and interacting via arguably the simplest possible interaction of steric hindrance has received little attention in the literature. The focus on point-like constituents largely stems from the challenges associated with a theoretical analysis of a system of interacting extended objects in terms of suitable coarse-grained variables, in the spirit of a tractable statistical mechanical description, and the lack of a general prescription to guide one with the identification of such variables. In this work, we aim to fill in these gaps, (i) by showing how a system of rigid extended objects interacting through steric hindrance may exhibit nontrivial long-time behavior, both static and dynamic, including a noise-induced phase transition, non-monotonic dependence of the field-induced velocity as a function of the field strength \textit{even in the absence of any quenched disorder}, and (ii) by adducing a remarkable and hitherto-unexplored coarse-grained description that allows to capture effectively the dynamics of the system and in particular the signatures  of the underlying phase transition.     

Understanding how extended objects move individually in space~\cite{aguilar2016review} as well as how a collection of them behave and organise themselves~\cite{popkin2016physics,feinerman2018physics} has been a topic of recent research \cite{chate2020dry}. However, study of the role of their shapes in determining the long-term behavior of the collection has mostly been limited to the case of elongated shapes inducing alignment~\cite{narayan2007long,gompper20202020,kudrolli2008swarming}. The fact that shapes can interlock and hence attain different functionality has largely been explored in the colloidal literature~\cite{sacanna2010lock,wang2012colloids}. The role of this interlocking in macroscopic systems has received little attention~\cite{ghosh2018geometric,penrose1957self}.  

A general extended object would consist of bounding surfaces that at some parts are locally concave while at  others are convex. This gives rise to the possibility of interlocking between locally convex and concave regions of these objects through compensation of curvature. However, even convex structures like truncated polyhedron can interlock through compensation of slopes (and not of curvature) of the engaging surfaces~\cite{siegmund2016manufacture}.

\begin{figure}[t]
		\centering
		\includegraphics[width=.9\linewidth]{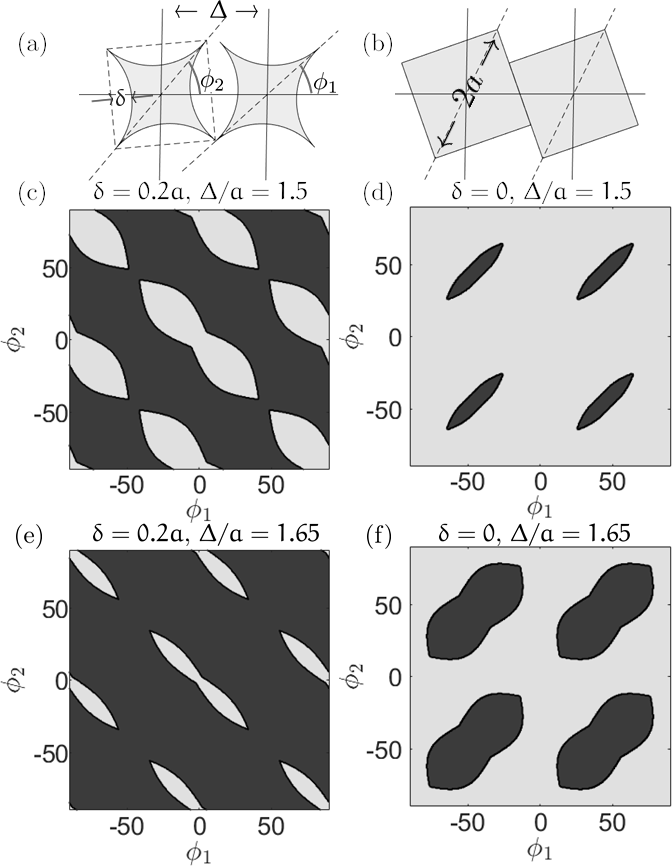}
		\caption{Two curved and flat squares of diameter $2a$ whose centers are separated by $\Delta$ is shown in panels (a) and (b), respectively. The angles $\phi_1 \in [-180^{\circ},180^{\circ})$ and $\phi_2 \in [-180^{\circ},180^{\circ})$ characterize the orientation of the two squares.  Each curved square is made by subtracting  four  equal circular segments of height $\delta$ from a square  of length $\sqrt{2}a$. Panels (c), (d), (e) and (f):  Accessible (dark gray)  and  inaccessible (light gray) regions  curved out of the configuration space $(\phi_1, \phi_2)$ of two curved squares. The boundary between the two regions corresponds to configurations in which the two squares touch each other.  	
		For two different values of $\Delta$, while (c) and (e) represent the case of $\delta=0.2a$, panels (d) and (f) represent flat squares ($\delta=0$).}
		\label{fig:Fig1n}
	\end{figure}

\section{1-dimensional periodic array of rigid polygons}
 
When a large number of extended objects, among which interlocking may happen, move in a common space, it is evidently of interest to ask about the configuration(s) that the system would settle to in the stationary state attained at long times under the dynamics of transition between the various possible configurations.  In order to explore this issue, we will consider in this work a representative system comprising a one-dimensional periodic array of $n > 1$ rigid polygons each of which is undergoing in presence of drive and noise a rotational motion in two dimensions about its fixed center under the constraint that no two polygons overlap (steric hindrance).  
All the polygons are identical in every respect.  The drive is the same for all the polygons, and rotates them individually in the same direction that we may take to be the clockwise direction without loss of generality. The noise is taken to be uncorrelated in time as well as between the different polygons. As representative examples of polygons in two dimensions, we will consider curved squares obtained by subtracting four equal circular segments of height $\delta >0$ from a square  of length $\sqrt{2}a$, see Fig.~\ref{fig:Fig1n}(a). In the limit of $\delta $ going to zero, the polygons are flat squares. By controlling the height of the circular segments, one may generate a variety of shapes.

In our model, we may already anticipate nontrivial physical effects on the basis of the fact that the constituent objects are extended. Let us first consider the no-noise situation. Referring to Fig.~\ref{fig:Fig1n}, panels (a) and (b), we realize that as the polygons are being rotated in the clockwise direction, there may come an instant of time at which neighbouring polygons get interlocked. In such a situation, further increase of the drive strength would only reinforce the interlocking. What noise is expected to do is to bring about unlocking, provided that the noise is strong enough. What is the physical implication of interlocking? It would lead to a state in which despite the drive, the polygons would not be rotating and hence, there will be no longitudinal current along the one-dimensional array; here the longitudinal current at a given time instant is defined as the angular velocity  averaged over neighbouring polygons. Interlocking will happen only when for a fixed height $\delta$ of the circular segments, the ratio $\Delta/a$ of the separation $\Delta$ between the centers of two consecutive polygons to the diameter $a$ of the polygons is small, see Fig.~\ref{fig:Fig1n}(a). As the ratio increases, the polygons get further apart. Beyond a critical value of this ratio, the polygons cannot interlock, and then there will be a non-zero current along the array at any drive strength. It is worthwhile to mention that the zero-current state could have been trivially reached if we had chosen the drive strength to be a quenched-disordered random variable different for different polygons, so that some of them rotate slower than others, and then obviously the slowest of all polygons would have led to long stretches of time over which neighboring polygons interlocked with this slowest polygon do not get to rotate. In this sense, the zero-current state if and when attained in our system would be a spontaneously-generated state emerging not owing to any disorder but solely from the steric hindrance between neighboring polygons. On the basis of the foregoing, we thus see an intricate and nontrivial interplay of the various dynamical parameters at work: the strength of the noise, the drive strength, the parameters $\delta$ and $\Delta/a$. It is certainly of interest to chart out the possible long-time dynamical scenario as one varies these parameters. It is evident that interlocking plays an essential role in generating nontrivial dynamics: had we taken objects that albeit extended do not interlock into one another (e.g., discs), one could not have generated a zero-current state.

Now, since each polygon has one degree of freedom (the angle $\phi \in [-180^{\circ},180^{\circ})$, see Fig.~\ref{fig:Fig1n}(a)), the dynamics takes place in the $n$-dimensional configuration space of the system. However, owing to the non-overlapping constraint, the dynamics would be restricted to only certain accessible regions of the configuration space. While the boundaries between the accessible and the inaccessible region correspond to the polygons having at least one point of contact, the interior of the accessible regions would correspond to no contact and hence unhindered motion of the polygons. As the configuration of the polygons changes with time, so does the number of contacts among them. Consequently, the dynamics of the system may be characterized by monitoring how the number of such contacts changes as a function of time. However, characterizing the dynamics by the number of contacts does not keep track of the identity of the polygons in contact. In this sense, the dynamics of contacts offers a coarse-grained lower-dimensional characterization of the dynamics of polygons. We may invoke a further coarse-graining by considering the number of contacts to be a binary variable  taking values $1$ or $0$ corresponding respectively to either presence or absence of contact. In this sense, we do not make any distinction between panels (a) and (b) of Fig.~\ref{fig:Fig1n} as regards number of contacts. Note that since we are dealing with those shapes that can interlock, the  boundary between the accessible and inaccessible regions will have locally concave and convex regions. As already mentioned above, we  do not consider disk-shaped objects, which evidently do not interlock. 

\subsection{ Configuration space for  two polygons}

Let us illustrate the aforementioned idea of dynamics of contacts with respect to Fig.~\ref{fig:Fig1n}(a). The accessible and the inaccessible region that is curved out of the configuration space ($\phi_1, \phi_2$) correspond respectively to the dark gray and the light gray region  in Fig.~\ref{fig:Fig1n}(c) -- (f), while the boundary between the two regions corresponds to configurations in which one polygon touches the other. 
 Figure~\ref{fig:Fig1n}, panels (c) and (e) correspond to curved squares ($\delta\neq 0$). As the ratio $\Delta/a$ is increased, the size of the inaccessible region grows and consequently the accessible space shrinks. A similar thing happens for flat squares ($\delta = 0$) as shown in Fig.~\ref{fig:Fig1n}, panels (d) and (f). A consequence of shrinking accessible region is a decrease in the longitudinal current of the system. For  the  shapes to rotate in a persistent manner, it is  essential that   the accessible region  forms  a continuous patch in the  ($\phi_1, \phi_2$) configuration space (see the dark gray patch in Fig. \ref{fig:Fig1n} (c) and (e))). In situations where this condition is not satisfied (isolated dark gray patches in Fig. \ref{fig:Fig1n} (d) and (f)),  the shapes cannot rotate persistently.

\subsection{The case of $n>2$ polygons: `Configuration space' $\{\phi_{1}\ldots \phi_{n}\}$  and   `Corner space' $\{D_{n-1}\ldots D_{0}\}$} 

For an periodic array of $n$ polygons, each performing a  rotation about a fixed point,  the angle $\phi_i(t)$ describes the trajectory  of the $i$-th polygon in time. The overall dynamics of the $n$ polygons happen in the  higher dimensional `configuration space' described by the coordinates  $\{\phi_{1}\ldots \phi_{n}\}$.  The corresponding `hypersurface'  embedded in the $n$-dimensional configuration space that is equivalent to the boundaries in Fig. \ref{fig:Fig1n}(c) -- (f)  would correspond to contacts whose value $N_c$ ranges from $1$ to $n$ in our coarse-grained description. Note that $N_c$ just counts the number of cases in a given configuration of polygons for which contacts have been made between neighboring polygons. These contacts define the dimensionality $D_m$ of the corners on the `hypersurface'  in the following way: $D_m=n d_{\rm f}-N_c$; here, $d_{\rm f}$ is the number of degrees of freedom of the individual shapes \cite{ghosh2018geometric}. In our example, since the polygons can only rotate in two dimensions about their fixed centres, we have $d_{\rm f}=1$. Thus, $D_{m}$ can take values as $\{D_{n-1}\ldots D_{0}\}$;~$D_n$ does not lie on the `hypersurface' and is hence excluded from the counting.  The coordinates $\{D_{n-1}\ldots D_{0}\}$ describe the `corner space'.

\subsection{ Markovian dynamics on space defined by  the dimesionality of the corners}

Corners of different dimensionalities (i.e, the $D_m$'s) can be thought of as states that are populated as the system of polygons under consideration evolves in time from one configuration to another. The dynamics of transition from one state to another can then be encoded in the form of a Markov chain generated by a  transition matrix whose elements give the probability of transition from one state to another in an elementary time step of the dynamics.  Specifically, the off-diagonal elements of the  transition matrix give the probability of transition between two different states, while the diagonal elements give the probability of dynamical self-loops whereby a state does not evolve in time. Invoking Markovianity may be justified on the basis of the fact that what a given state would evolve to would depend just on itself and not on how the given state is arrived at dynamically. We have discussed later in the text a numerical check of Markovianity of the coarse-grained dynamics. 

It is pertinent to state right at the outset our main results that concern the stationary state of the system attained at long times. 
We construct from the motion of the polygons the dynamics of transition (in particular, the probability of transition) between the corner states and thereby obtain the stationary state of the corner dynamics. We establish that the stationary-state dynamics of the corner states is Markovian in nature.  The stationary state of the dynamics in the corner space is found to coincide with the one obtained on feeding the constructed probability of transition between the corner states into the Markov chain model of corner dynamics and solving it to obtain the stationary state.  The stationary state obtained from analyzing the Markov chain implies an inhomogeneous probability measure over the corner states, whereby certain specific states are preferred over others. Remarkably, for large $n$, the relative ordering of the preference varies non-monotonically with the dimensionality of the corners and hence with the number of constraints, implying thereby that a moderately-constrained state is the most preferred one. Had the probability measure been homogeneous, which would have been the case in the absence of drive or contacts, one would have had a monotonic dependence on the number of constraints, with the least-constrained state being the most preferred state. A particularly remarkable revelation of our study is the fact that the stationary-state probability measure of corner states for varying $n$ shows distinct scaling behavior, implying thereby that the state space for different $n$ is self-similar. A practical utility of self-similarity is that one can extract the long-time state of a system for large $n$ by studying a much smaller system.

Our next salient result concerns the average angular velocity  of polygons in the stationary state of the system. We show that at a fixed noise strength, the average angular velocity shows a striking non-monotonic dependence on the strength of the drive. On the other hand, this velocity when considered at a fixed drive and as a function of the noise strength shows an abrupt increase in its value beyond a critical noise strength. This effect conforms to what we had anticipated earlier in the paper that interlocking does not allow the velocity to grow unless the noise is of critical strength. Then it succeeds in unlocking the polygons and causes the velocity to grow. The behavior of the velocity as a function of the noise strength resembles a noise-induced phase transition. Based on our gathered wisdom regarding phase transitions, one would expect a different behavior of the relaxation times across the critical point. Indeed, on analyzing the gap between the leading and the sub-leading eigenvalue of the transition matrix of the Markov chain model, we find that the gap that dictates the timescale of relaxation of the Markov dynamics shows a marked change in its behavior close to the critical noise strength. Although not evidently true, such a match is a demonstration that our coarse-grained description in terms of corners on the hypersurface captures quite remarkably the stationary-state behavior of the polygon dynamics.

\begin{figure}[t]
         \centering
     \includegraphics[width=1\linewidth]{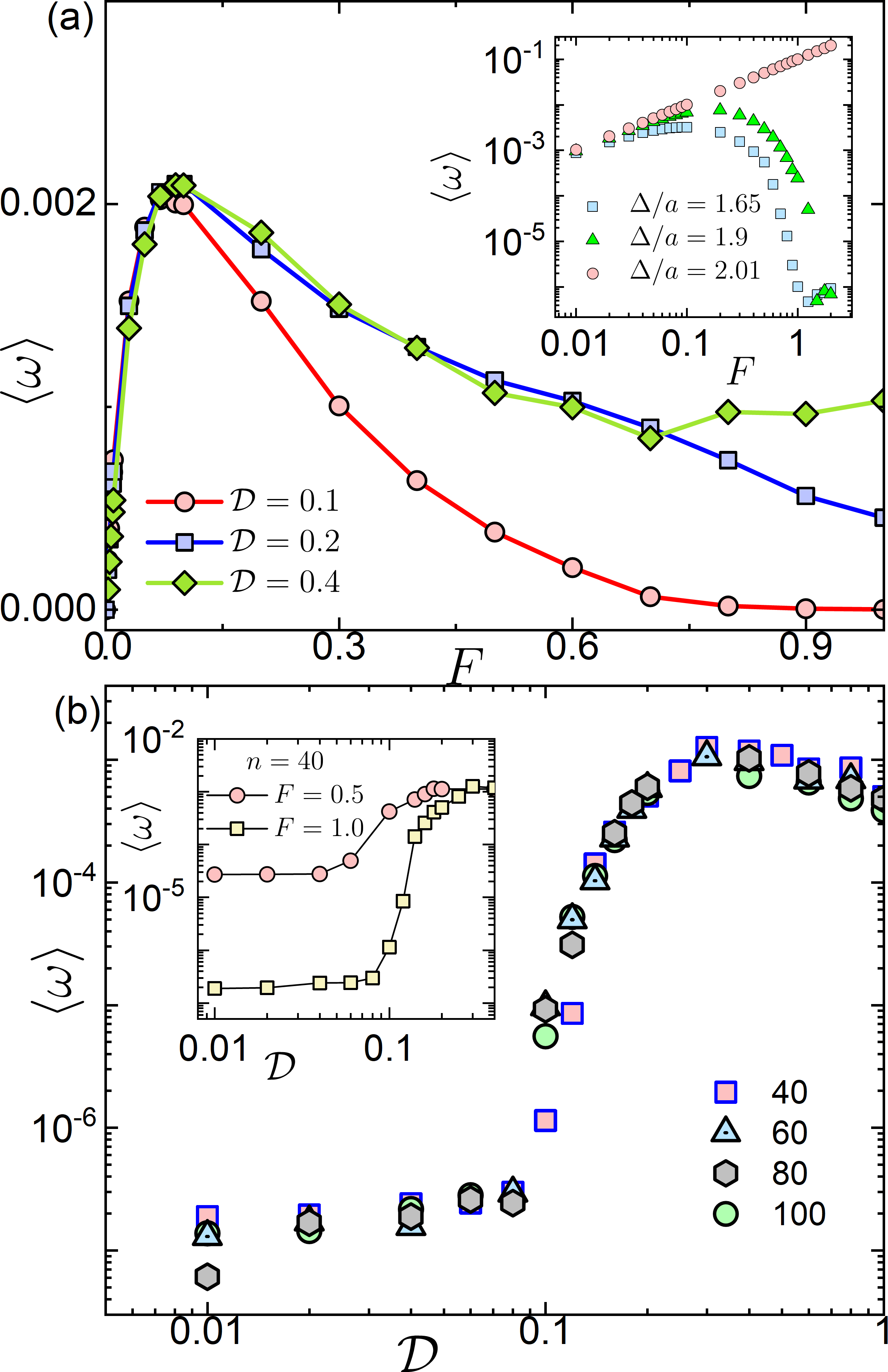}
         \caption{         (a) Average angular velocity $\langle \omega \rangle$ as a function of the  drive strength $F$ in the stationary state of the system for three different values of the noise strength $\mathcal{D}$ for a chain of $n=40$ curved squares, where the squares are separated by a distance $\Delta=1.65a$. The inset shows the variation of $\langle \omega \rangle$ with $F$ for  representative values of the ratio $\Delta/a$. (b) Variation of $\langle \omega \rangle$ as a function of noise strength ${\cal D}$ at a fixed bias $F=1.0$. The inset shows the   variation of $\langle \omega \rangle$  with the noise strength $\mathcal{D}$ for  representative values of the  drive strength $F$ for $n=40$.}
        \label{fig:current_drive}
     \end{figure}	
                
\section{Methods}
          
We now turn to a derivation of our results.
For our one-dimensional periodic array of curved polygons, the dynamics involves noisy rotation of each polygon in two dimensions  about its fixed centre in presence of a drive. In other words, each polygon performs a driven-diffusive motion in the space of its orientation angle $\phi \in [-180^{\circ},180^{\circ})$. The rotational motion of the polygons during a small time interval between $t$ and $t+\Delta t$ corresponds to update of the angle $\phi_i;~i=1,2,\ldots,n$ as
\begin{align}
     \phi_i(t+\Delta t)=\phi_i(t)+{\cal D}\eta_i \sqrt{\Delta t} + F \Delta t.
	 \end{align}
	 Here, 	$\eta_i$ is a mechanical noise that causes independent rotation of the squares, and $F$ is the drive or the bias. The parameter ${\cal D}$ characterizes the strength of the noise term. We assume $\eta_i$ to be a Gaussian, white noise, namely,  with zero mean and correlations in time given by  $\langle \eta_i(t)\eta_j(t)\rangle=\delta_{ij}$ and $\langle \eta_i(t)\eta_j(t')\rangle=0$ for $t \ne t'$. Note that the noise acting on different polygons are completely uncorrelated.  In order to implement the non-overlapping constraint between neighbouring polygons in numerical experiments that we perform with the polygons, we associate with each polygon a set of $2d$ points that are contained inside it. Let $Q_i(t)$ denote this set for the $i$-th polygon at time $t$; $i=1,2,\ldots,n$.  At a given time instant, the neighbouring polygons  are considered to be in contact if their edges are within a predetermined distance $\varepsilon$. We consider noisy instead of deterministic dynamics, since in absence of noise, the biased system would for high bias values get jammed due to interlocking and the dynamics would remain stuck in such a configuration for all subsequent times. Presence of noise allows the system to escape from such jammed configurations.
	 
Corresponding to the rotational motion of the polygons, all the elements of $Q_i$ undergo the same rotational transformation: $Q_i(t+\Delta t) = R(\Delta\phi_i) Q_i(t)$. Here, $R(\Delta\phi_i)$ is the $2d$ rotation matrix, and $\Delta\phi_i \equiv \phi_i(t)-\phi_i(t+\Delta t)$ is the incremental change in the orientation $\phi$ in time $\Delta t$. All  the  $\phi_i$'s are updated sequentially  at each  simulation time step subject to the constraint that the intersection of no two sets $Q_i$ is a null set, i.e., $Q_i \cap Q_j =\emptyset~\forall~i,j$. 
     For cases where the update results in one shape overlapping with the other, i.e., $Q_i \cap Q_j\ne \emptyset$, the proposed dynamical move is rejected in numerical experiments. In such situations, the shape continues to be in its previous state, i.e., $Q_i  (t+\Delta t) =Q_i  (t)$. Two  neighbouring shapes  with set of points $Q_i$ and $Q_{i\pm 1}$ are said to be in contact if they are within a neighbourhood of each other, i.e.,  if the pairwise Euclidean distance matrix between them has values smaller than $\varepsilon$. 
     
\section{Results and Discussion}
     
 \subsection{Dynamics in the configuration space $\{\phi_1, \phi_2, \ldots \phi_n \}$ }
     
In the absence of bias, $F=0$,  a clockwise move is as likely as an anticlockwise move, and the  resulting dynamics in $\phi$ is diffusive. However, if  $F\ne 0$, the resulting dynamics is a driven diffusive motion, whereby the polygons move preferentially in  either the anti-clockwise  or the clockwise direction (in clockwise direction, in the convention we have adopted). The bias $F$ has two effects on the overall dynamics: (i) it generates a rotational  drift,  and (ii) it enforces the  interlocking of the shapes.  We find that the average angular velocity  $\langle \omega \rangle$  varies non-monotonically with the strength $F$ of the bias: it increases initially with the bias; however, larger bias values enforce interlocking and the average velocity drops to zero. Beyond a critical bias $F^{c}$, the  mobility of  the system  becomes negative. The variation of the average angular velocity  $\langle \omega \rangle= \langle \phi_i(t)-\phi_{i}(t+\Delta t) \rangle _{(t, i)}$  with the strength of the bias  $F$   for three representative values of noise strength $\mathcal{D}$ is shown in Fig. \ref{fig:current_drive}(a). The inset to the figure shows the variation in the angular velocity  as a function of the bias  $F$  for different values of the parameter $\Delta/a$.  We find that when we have $\Delta/a > 2$, the polygons never  touch each other and the average angular velocity  increases linearly with the drive. For values of $\Delta/a \le 2$, the  angular velocity changes non-monotonically with  $F$. However, the magnitude of $\langle \omega \rangle$ for a given value of  drive $F$ decreases with decrease in $\Delta/a$.  Similar decrease in $\langle \omega \rangle$  for a given drive is observed for   decreasing   $\delta$ that is measure of the  extent of curvilinearity.
	
It is worthwhile to mention that similar nonmonotonic dependence of the velocity on field strength for random walks in presence of a bias has been observed previously, see Ref. \cite{white1984field}. However, an important difference with respect to our work is that in \cite{white1984field}, the random walk takes place on a disordered lattice, e.g., on a percolation cluster. The appearance of a critical bias above which the velocity becomes zero is attributed to the presence of so-called backbends in the lattice escaping from which requires motion opposite to the bias direction and which consequently becomes less and less probable with increasing bias. Similar effects have recently been observed in colloidal transport process in presence of obstacles \cite{eichhorn2010negative,reichhardt2017negative}.  However, in our case, there is no disorder in the system. 

\begin{figure}[t]
         \centering
        \includegraphics[width=1\linewidth]{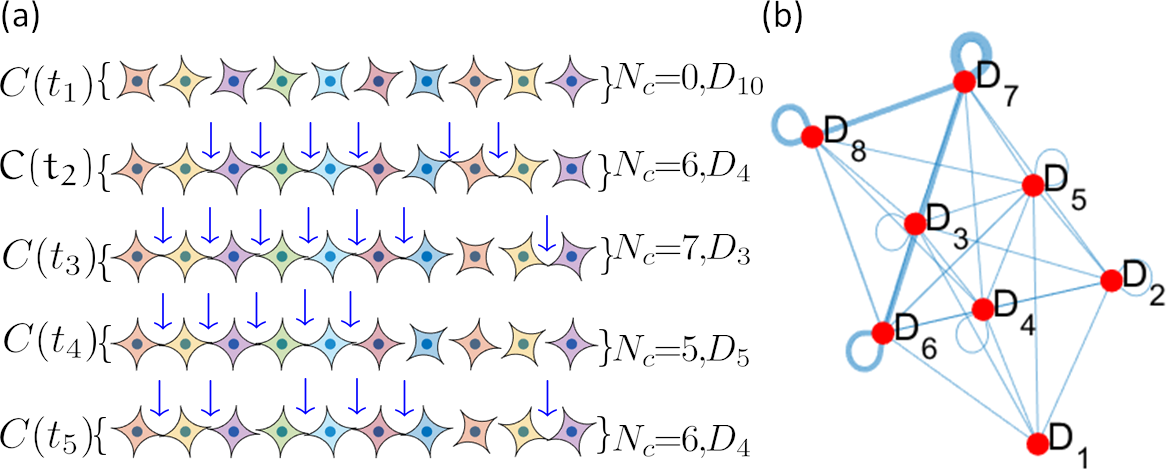}
       \caption{(a) Representative  configurations $C(t)$ for our model of one-dimensional array of $n=10$ curved squares rotating about their fixed centers, see text.  The number of constraints $N_{c}$ and the dimensionality $D_m$ associated with each configuration $C(t)$ are mentioned along with it. The arrows mark the spatial position at which the squares touch each other. A given dimensionality of the corner may correspond to same number of arrows that are however arranged differently, see the case for $D_4$. (b) The graph represents an example of possible transitions between various states $D_{9} \ldots D_{1}$ in the system, where the nodes denote the states and the arrows denote the direction of transition. 
         }
         \label{fig:Fig3}
     \end{figure}

The strength of the noise $\mathcal{D}$ plays an important role in determining the dynamics of the system. As discussed in Fig.~\ref{fig:current_drive}(a), with increasing drive $F$, the curved polygons interlock and the average angular velocity $\langle \omega \rangle$ goes to zero. This  cessation of the persistent rotational motion is most pronounced when the  strength of the noise is weak. With increasing noise strength, two types of dynamics begin to compete: (i)  interlocking that is enforced by the  drive $F$,  and (ii)  unlocking that is facilitated by the noise whose strength is $\mathcal{D}$. Beyond a critical  noise strength,  the  unlocking phenomenon becomes more dominant. In this parameter regime,    the angular velocity $\langle \omega \rangle$ begins to increase as a function of  the noise strength $\mathcal{D}$. This noise-induced activity is shown in Fig.~\ref{fig:current_drive}(b), where we plot the variation of the angular velocity $\langle \omega \rangle$  as a function of the  noise strength $\mathcal{D}$  for $F=1.0$. This variation is plotted for various values of $n$.  This  critical noise strength decreases with  lowering of the drive strength $F$. This is shown in the inset to Fig.\ref{fig:current_drive} (b). 

\begin{figure}[b]
    \centering
   \includegraphics[width=1\linewidth]{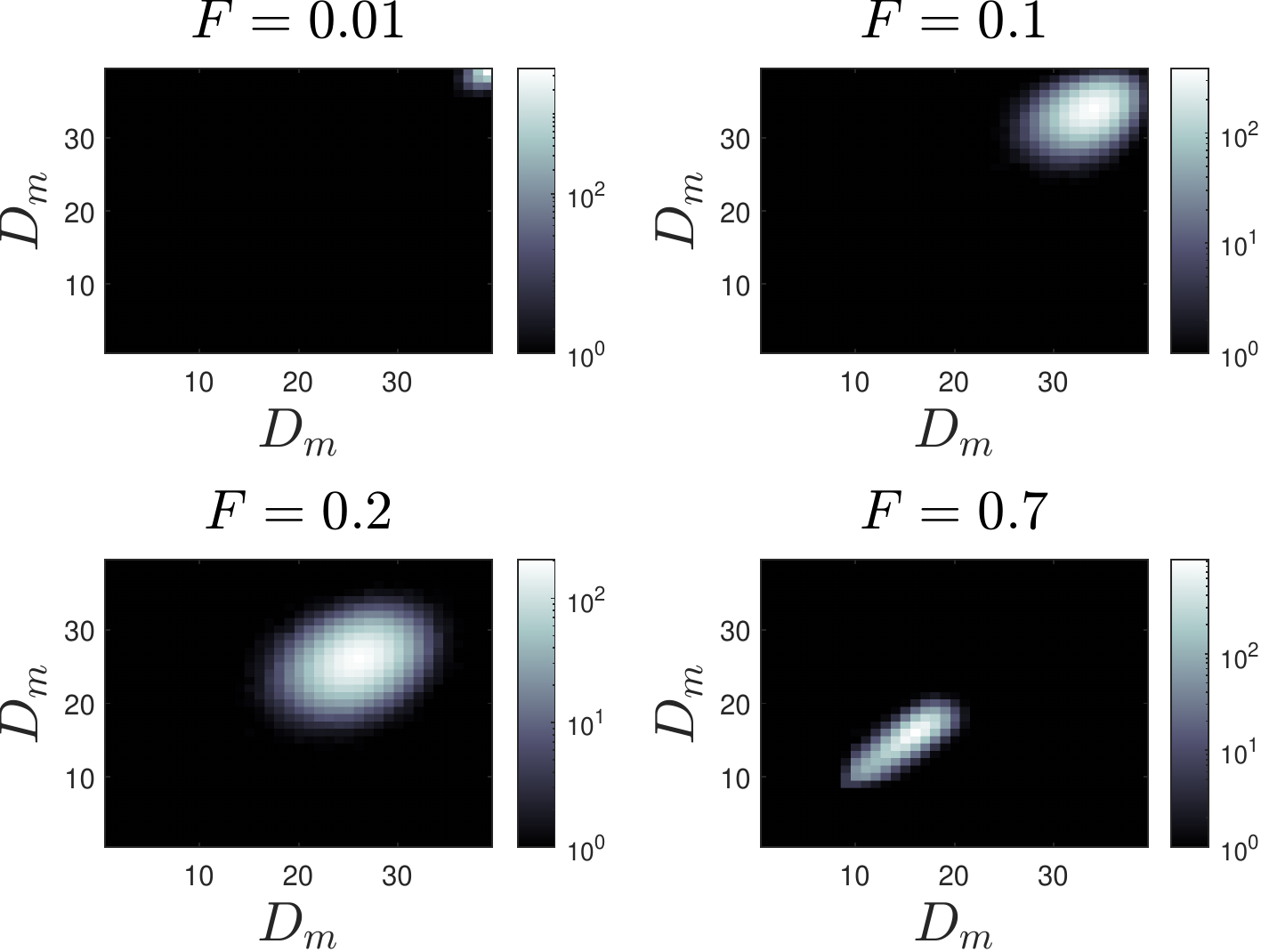}
    \caption{Shows the state transition matrix $T$,  for representative values of drive strength  $F$ . Here $n=40$ and $\mathcal{D}=0.1$. }
    \label{fig:transition_matrix}
\end{figure}
     
     \begin{figure}[t]
    \centering
   \includegraphics[width=1\linewidth]{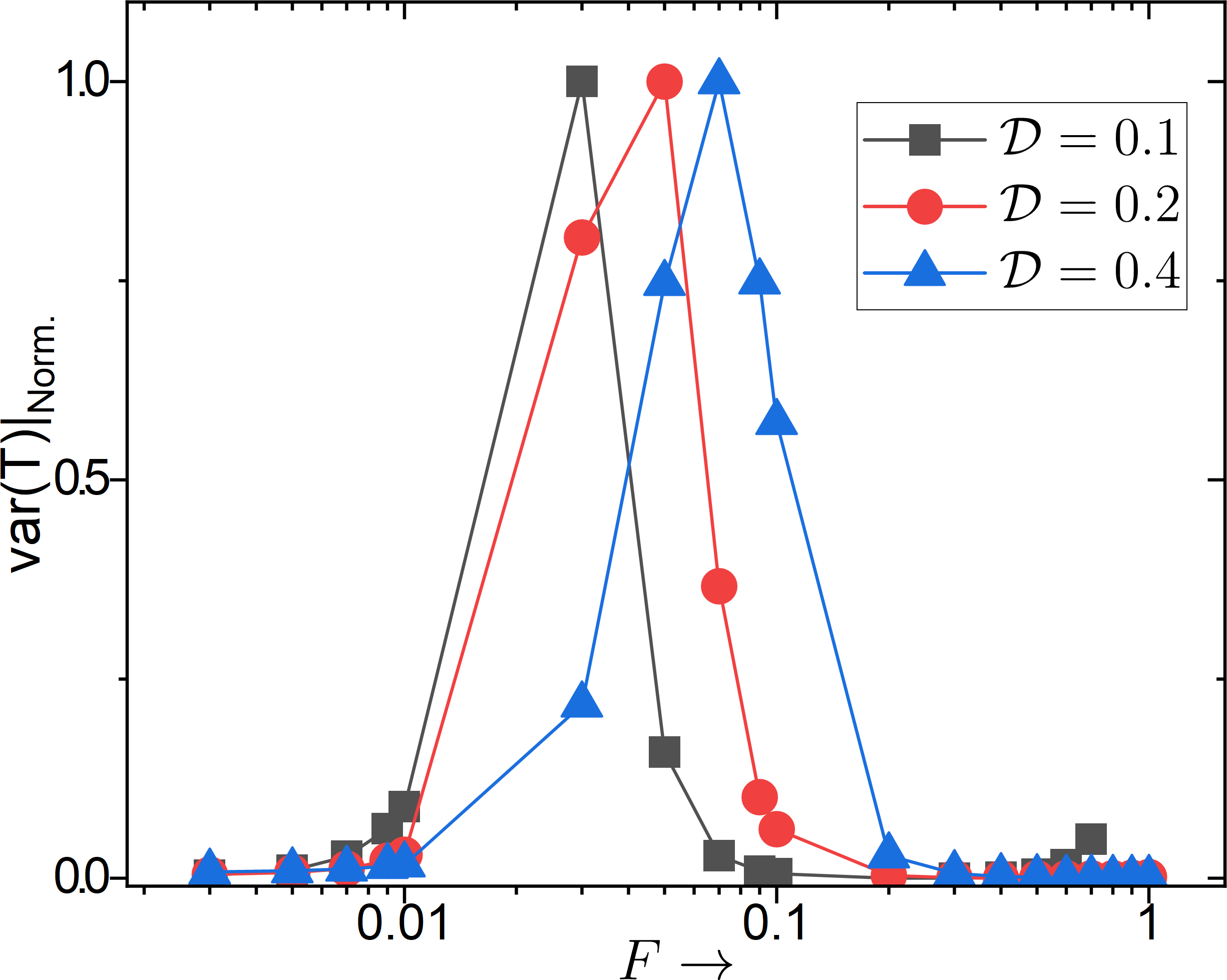}
    \caption{The figure plots the   variance $ \mathrm{var}(T)|_{\mathrm{Norm.}} $ of the  entries of the  state transition  matrix $T$ as a function of $F$  for  three different values of the  noise strength $\mathcal{D}$ , with each  curve  normalised by its  maximum value.}
    \label{fig:transition_matrix_var}
\end{figure}

\begin{figure}[b]
    \centering
   \includegraphics[width=.8\linewidth]{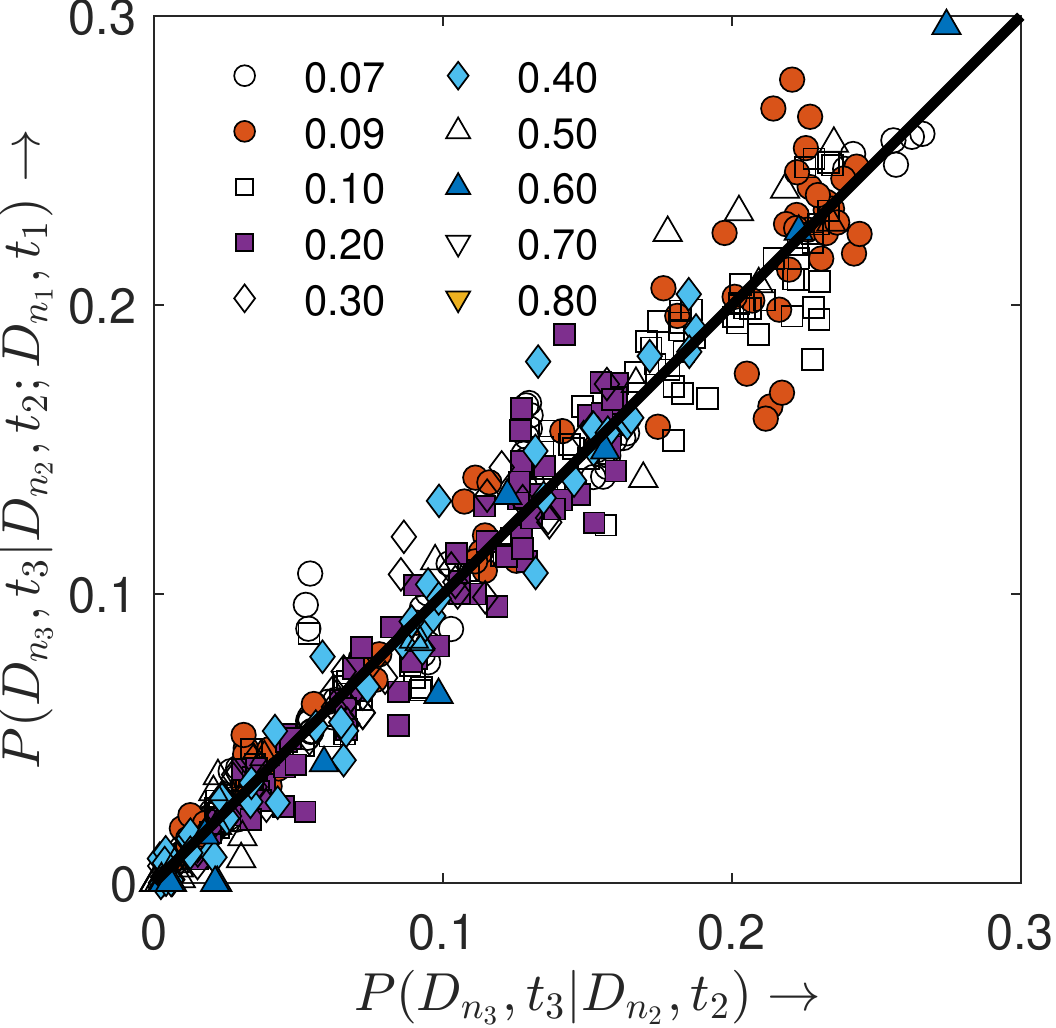}
     \caption{The figure checks the Markovianity of our coarse-grained dynamics: We find that $P(D_{n_3},t_3|D_{n_2},t_2;D_{n_1},t_1) \approx P(D_{n_3},t_3|D_{n_2},t_2)$ for $n=40$,  $\mathcal{D}=0.1$ and representative values of  drive strength $F$ (marked in the figure).  Numerically, we have obtained similar results for several representative values of $D_{n_1}$, $D_{n_2}$ and $D_{n_3}$ and also of $t_1$, $t_2$ and $t_3$.  The solid line represents the condition $P(D_{n_3},t_3|D_{n_2},t_2;D_{n_1},t_1) = P(D_{n_3},t_3|D_{n_2},t_2)$.}
    \label{fig:markov}
\end{figure}

\subsection{Dynamics in the corner space $\{D_1, D_2, \ldots D_n \}$ }
 	   
We now discuss our coarse-grained model. Representative configurations $C(t)$  are shown in Fig. \ref{fig:Fig3}(a).    The number of constraints $N_{c}$ and the dimensionality $D_m$ associated with each configuration $C(t)$ are mentioned  along with it. The arrows mark the spatial position at which the  squares touch each other. In order to analyze the dynamics of the rotating squares in terms of $D_m$'s, we now introduce a discrete-time Markov chain whose state space is constituted by the $D_m$'s and in which transitions between the various states are encoded in a transition matrix $P$. For a system of $n$ squares, the transition matrix is a $n \times n$ matrix such that its $mm'$-th element gives the probability of transition from state $D_m$ to state $D_{m'}$. Contrary to the usual treatments of stochastic processes wherein the transition matrix is given as part of modelling of the dynamics, here we construct the  state transition matrix from observing the dynamical trajectories of the $D_m$'s in simulations. Specifically, considering a dynamical trajectory of a sufficiently long duration, the transition probability from state $D_m$ to $D_{m'}$ is estimated as the fraction of the total time at which one observes between subsequent time instances a transition $D_m \to D_{m'}$. This gives the state transition matrix $T$. In order to obtain the transition matrix $P$, we normalize the elements of $T$ in a column such that they add up to unity. The graph in Fig. \ref{fig:Fig3}(b)  depicts the transitions between the states $D_1,D_2 \ldots D_9 $ for a linear array of $10$ squares. The nodes in the graph represent the various states $D_m $ and the arrows denote the direction of transitions. 
 	   
\subsection{Inhomogeneous  nature of the transition matrix $T$} 
 	   
Figure \ref{fig:transition_matrix} shows the state transition matrix $T$, for representative values of drive strength  $F$ for $n=40$. The corresponding dynamics   in the configuration space is shown in Fig.\ref{fig:current_drive} (a). The relative magnitudes of the different transitions are represented by the grey scale shown alongside. The correctness of our estimation of the transition probabilities can only be judged \textit{a posteriori} from, e.g., a comparison of the numerical result for the stationary-state distribution of $D_m$'s with the one obtained from a theoretical analysis that we detail below. 
 	   
The behavior of the average current $\langle \omega \rangle$ in the configuration space (Fig.~\ref{fig:current_drive}) is  reflected in the   degree of inhomogeneity of the state transition matrix $T$.  The degree is quantified by measuring the variance of the entries of the matrix $T$. Figure \ref{fig:transition_matrix_var} shows the plot of the   variance $ \mathrm{var}(T)|_{\mathrm{Norm.}} $ of the  entries of $T$ as a function of the drive  $F$ for representative values of  the noise strength $\mathcal{D}$, in which each  curve is normalised by its  maximum value. For small values of the drive, the system encounters fewer steric hindrances, as a result of which the number of points on the hypersurface is small, and  in this phase, $\langle \omega \rangle$  grows with $F$. As the drive increases, the  number of points on the hypersurface increases, and the system becomes  more constrained. In this phase, the matrix $T$  becomes increasingly more inhomogeneous, and the variance of its entries grows with the drive. Increasing constraints hinder the motion of the squares,  and hence, in this phase,  $\langle \omega \rangle $  decreases with $F$.  With still increasing drive, additional states get populated, and this finally begins to make the matrix $T$ appear more homogeneous. However, these additional constraints drive the system into an over-constrained state where adding  extra constraints via increasing  $F$ have little effects on $\langle \omega \rangle$.

 \subsection{ Markovian dynamics on the corner space}
 	    
\subsubsection{A check of Markovianity}
Given the stationary-state time series data for  $D_m$'s, we check in the following manner that the dynamics of our coarse-grained model is Markovian. Given three time instances $t_1<t_2<t_3$, let us define $P(D_{n_3},t_3|D_{n_2},t_2;D_{n_1},t_1)$ as the conditional probability for the system to be in state $D_{n_3}$ at time $t_3$ while conditioned on having been in state $D_{n_2}$ at time $t_2$ and in state $D_{n_1}$ at time $t_1$. On the other hand, we define $P(D_{n_3},t_3|D_{n_2},t_2)$ as the probability for the system to be in state $D_{n_3}$ at time $t_3$ while conditioned on having been in state $D_{n_2}$ at time $t_2$. A check of Markovianity of the given time series is to confirm that one has~\cite{tabar2019}
 	   $P(D_{n_3},t_3|D_{n_2},t_2;D_{n_1},t_1)=P(D_{n_3},t_3|D_{n_2},t_2)$.
      We have checked this numerically for several representative values of $D_{n_1}$, $D_{n_2}$ and $D_{n_3}$ and also of $t_1$, $t_2$ and $t_3$, see Fig.~\ref{fig:markov}.

\subsubsection{A comparison between direct numerical simulation and stationary state distribution }

Following~\cite{vankampen}, we now analyze the Markov chain introduced in the foregoing. As already discussed, the states of the chain are the corner $D_m$'s, and the transition matrix $P$ encodes the information on the probability of transition between states. For a system of $n$ squares, for which $P$ is an $n\times n$ matrix, the column sum of $P$ is unity since a transition in an elementary time step should take the system to either a different state or to itself. The former fact is encoded in saying that $P$ has a left eigenvector with eigenvalue unity given by $\langle 1_L|=(1 1 \ldots 1):~\langle 1_L|P=\langle 1_L|$. Let the $(n-1)$-dimensional vector $|\pi\rangle$, which we call the probability vector, be such that its $m$-th element $\pi_{D_m}$ gives the probability for the system to be in state $D_m$. Being a probability vector, the sum of all entries in $|\pi\rangle$ is unity, i.e., $\langle 1_L|\pi\rangle=1$. It follows from the definition of the transition matrix that $|\pi\rangle^{(1)}=P|\pi\rangle^{(0)},$ where $|\pi\rangle^{(\alpha)};~~\alpha=0,1,2,\ldots$ denotes the probability vector at the $\alpha$-th time step. We have $| \pi \rangle^{(\alpha)} = (P)^{\alpha}|\pi\rangle^{(0)}$.

\begin{figure}[t]
       \includegraphics[width=1\linewidth]{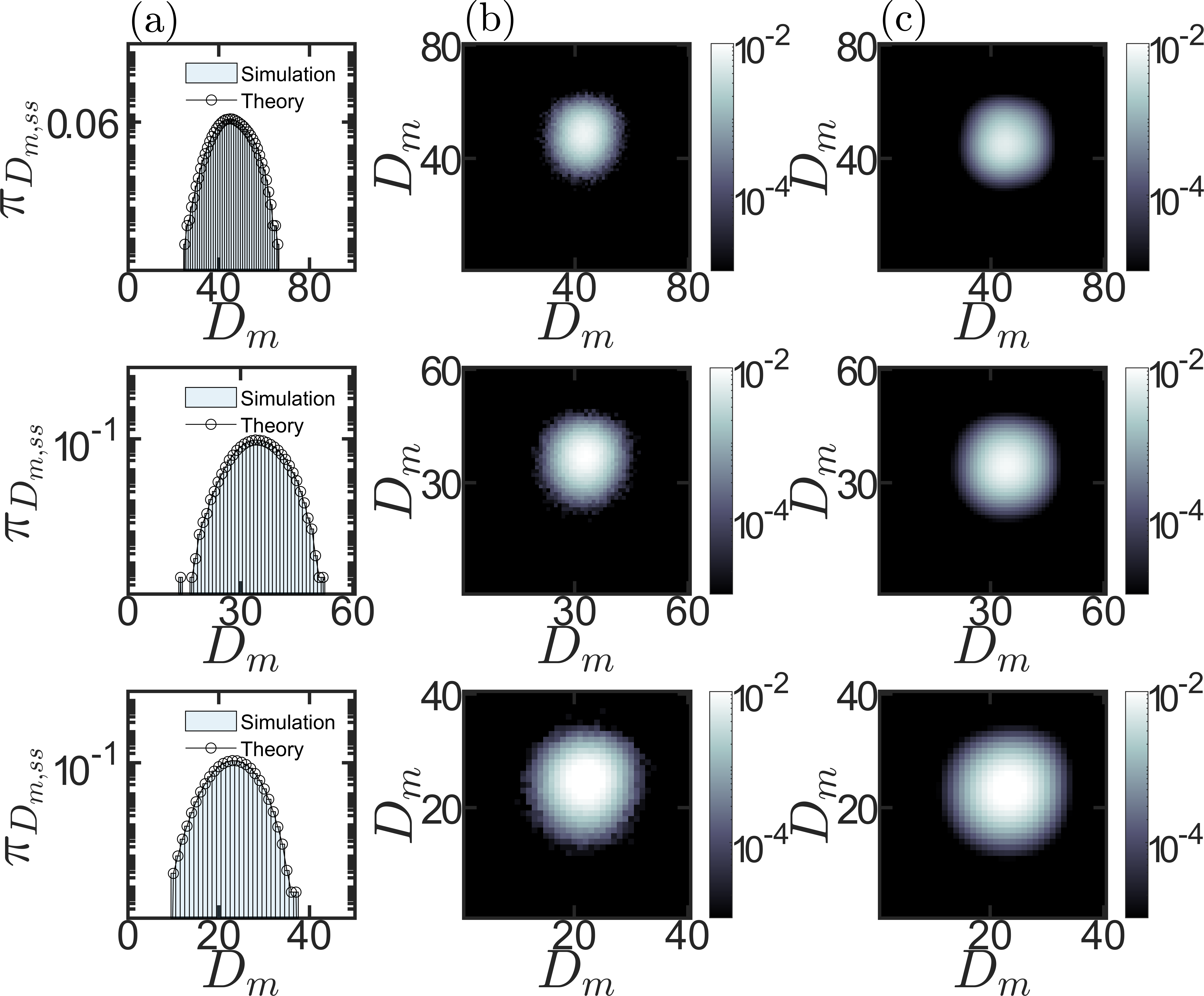}
       \caption{ (a) A comparison of the stationary-state distribution $\pi_{D_m}$ as obtained from direct simulation of the dynamics of the squares and from our theoretical analysis, ~Eq.~(\ref{eq:steady-state}), performed on the contact space; here, $F =1$ and $\mathcal{D}=0.4$. Note that $\pi_{D_m}$ shows a peak at $D_m=D_{\rm max}$. Panels (b) and (c) compare for $\tau=5000$ the stationary correlation $C_{D_{m}D_{m'}}(\tau)$ from direct simulations and theoretical analysis (Eq.~(\ref{eq:corr})), respectively. The top, middle and bottom panels correspond to $n=80$, $n=60$, and $n=40$, respectively. All results show an excellent agreement between theory and simulations
         }
         \label{fig:Fig4}
     \end{figure}
          	    
Corresponding to the dynamics initiated with the system being in state $D_{m_0}$, the initial probability vector has $\pi_{D_m}=\delta_{mm_0}$. A stationary state (ss) is characterized by a stationary probability vector $|\pi\rangle_{{\rm ss}}$ satisfying $|\pi\rangle_{{\rm ss}} =P|\pi\rangle_{{\rm ss}}$, i.e., $|\pi\rangle_{{\rm ss}}$ is given by the right eigenvector $|1_{R}\rangle$ of $P$ with eigenvalue unity:
\begin{align}
|\pi\rangle_{{\rm ss}}=\frac{|1_{R}\rangle}{\langle 1_L|1_R\rangle}.
\label{eq:steady-state}
\end{align}
The largest eigenvalue $\lambda_{1}$ of a transition matrix
is unity; the other eigenvalues $\lambda_{2}>\ldots>\lambda_{(n-1)}$
are less than unity~\cite{vankampen}. 
 
Let us define a correlation function $C_{D_mD_{m'}}(\tau)$ as the joint probability
in the stationary state for the system to be in state $D_m$ at a certain
time and in state $D_{m'}$ after a time lag $\tau$. We introduce
an indicator variable $n_{D_m}$ such that $n_{D_m}=1$ if the
system is in state $D_m$ and is zero otherwise. It follows by
definition that $(\overline{n_{D_m}})_{{\rm ss}}=\pi_{D_m,{\rm ss}}$,
where the overline denotes averaging in the stationary state. We may
define $
(\overline{n_{D_m}})_{{\rm ss}}=\langle1_{L}|n_{D_m}|\pi\rangle_{\rm ss},
$
 wherein the operator $n_{D_m}$ acting on $|\pi\rangle_{\rm ss}$ is defined
to yield $
n_{D_m}|\pi\rangle_{\rm ss}=[0 0 0 \ldots 
\pi_{D_m,{\rm ss}}\ldots 0]^T$, where $T$ denotes transpose. 
It follows that $\langle1_{L}|n_{D_m}|\pi\rangle_{\rm ss}=\pi_{D_m,{\rm ss}},$
as required. We have $C_{D_mD_{m'}}(\tau)=\langle1_{L}|n_{D_{m'}}(P)^{\tau}n_{D_m}|\pi\rangle_{\rm ss}$, yielding
\begin{align}
C_{D_mD_{m'}}(\tau)=\sum_{k=1}^{n-1}\frac{(\lambda_{k})^{\tau}}{\langle k_L|k_R\rangle}\langle1_{L}|n_{D_{m'}}|k_{R}\rangle\langle k_{L}|n_{D_m}|\pi\rangle_{\rm ss}.
\label{eq:corr}
\end{align}
Here, we have used  $P^\tau=\sum_{k=1}^{n-1}|k_{R}\rangle\langle k_{L}|\lambda_k^\tau/\langle k_L|k_R\rangle$, wherein $\langle k_L|$ and $|k_R\rangle$ are respectively the (unnormalized) left and right eigenvectors of $P$ with eigenvalue $\lambda_k$. In the above expression, $n_{D_{m'}}|k_R\rangle$ yields a $(n-1)$-dimensional column vector all of whose elements are zero except for its $m'$-th element being equal to the $m'$-th element of $|k_R\rangle$. 
As $\tau \to\infty$, since all the eigenvalues
excepting the largest one are smaller than unity, $
C_{D_mD_{m'}}(\tau\to\infty)=\langle1_{L}|n_{D_{m'}}|\pi\rangle_{\rm ss}\langle1_{L}|n_{m}|\pi\rangle_{\rm ss}=(\overline{n_{D_m}})_{{\rm ss}}(\overline{n_{D_{m'}}})_{{\rm ss}}$,
where we have used Eq.~(\ref{eq:steady-state}).
\begin{figure}[t]
         \centering
         \includegraphics[width=1\linewidth]{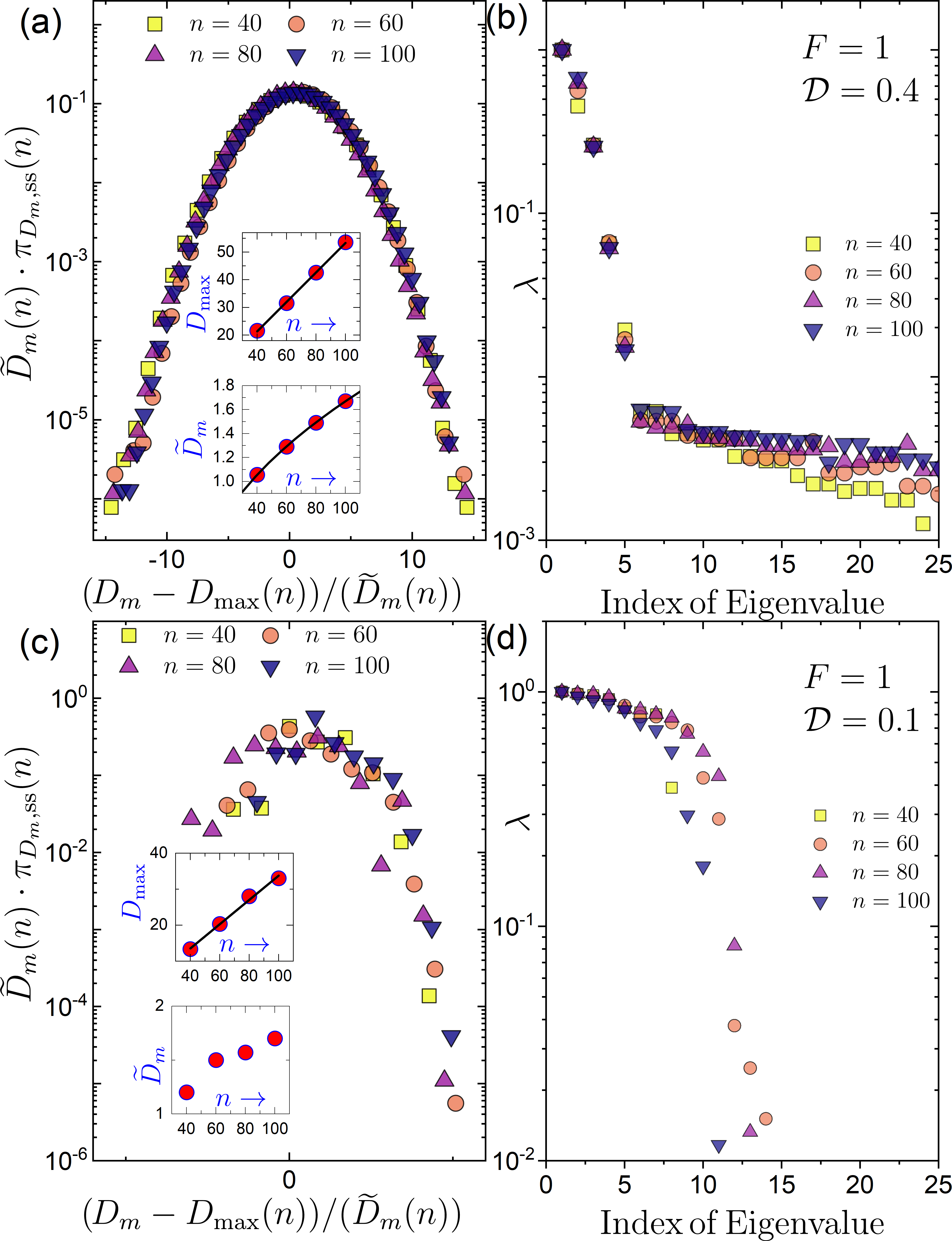}
         \caption{Scaling according to Eq.~(\ref{eq:scaling}) of the stationary  distribution $\pi_{D_m,{\rm ss}}$ for different $n$ are plotted in (a) for $F=0.1$ and $F =1.0$ and $\mathcal{D}=0.4$ and (c) for $F=0.1$ and $F =1.0$ and $\mathcal{D}=0.1$.  The insets to (a) and (c) show  the corresponding variations  in  $D_{\rm max}$ and   $\widetilde{D}_m$. For the inset in (a)          that $D_{\rm max} \propto n$  and   $\widetilde{D}_m \propto \sqrt{n}$.  However, for  the inset in (b) while $D_{\rm max} \propto n$, we find  $\widetilde{D}_m$  to only weakly depend on $n$.  The corresponding eigenvalue   spectrum of the  transition matrix $P$ as a function of the index  of the eigenvalue, for drive $F =1.0$ and $\mathcal{D}=0.4$ and $F =1.0$ and $\mathcal{D}=0.1$ is plotted in (b) and (d), respectively.}
         \label{fig:Fig5}
     \end{figure}

Figure~\ref{fig:Fig4}(a) compares the stationary-state distribution $\pi_{D_m,{\rm ss}}$ as obtained from direct simulation of the dynamics  of the  squares and from our theoretical analysis, ~Eq.~(\ref{eq:steady-state}), performed on the  contact space, demonstrating an excellent agreement.  This shows that the Markovian dynamics associated with the contact space captures the  dynamics of the higher-dimensional configuration space 
$\{\phi_1,\phi_2,\ldots,\phi_{n} \}$
of squares, thereby providing an effective lower-dimensional coarse-grained description of the problem. 
The data clearly suggest that the stationary-state probability of states is inhomogeneous, and consequently that the system is most likely to be found at a moderately constrained state, corresponding to the peak in the stationary-state distribution at $D_m=D_{\rm max}$. In Fig.~\ref{fig:Fig4}(b), we compare the stationary correlation $C_{D_{m}D_{m'}}(\tau)$ as obtained from direct simulations and from our theoretical analysis (Eq.~(\ref{eq:corr})), again displaying an excellent agreement. 

\subsection{Scaling properties of the stationary distribution $\pi_{D_m,{\rm ss}}$}

As shown in Fig.~\ref{fig:Fig5}(a), the stationary distribution $\pi_{D_m,{\rm ss}}$ for different $n$ shows a remarkable scaling behavior: 
\begin{align}
    \pi_{D_m,{\rm ss}}(n)=\frac{1}{\widetilde{D}_m(n)}F\left(\frac{D_m-D_{\rm max}(n)}{\widetilde{D}_m(n)}\right),
    \label{eq:scaling}
\end{align} 
where $F(x)$ is the scaling function, and  the parameter $\widetilde{D}_m(n)$ sets the scale of fluctuations of $D_m$ about $D_{\rm max}(n)$. We find that $\widetilde{D}_m(n)$ scales as $\sqrt{n}$  while the peak of the distribution $D_{\rm max}(n)$ scales linearly with  the number of polygons $n$.  Such a behavior would have been expected if the motion of the polygons were uncorrelated; the fact that  we see a similar behavior in our case implies weak correlation among the motion of the different polygons, which is a manifestation of the fact that Fig.~\ref{fig:Fig5}(a) corresponds to the situation in which the average angular velocity $\langle \omega \rangle$ is non-zero  ($F=0.1$ and $\mathcal{D}=0.4$). This case may be contrasted with the plot shown in Fig.~\ref{fig:Fig5}(c) for which we have $\langle \omega \rangle $ is almost zero  ($F=0.1$ and $\mathcal{D}=0.1$), and consequently, the concomitant strong correlations between the motion of the polygons reflect in a different scaling of $\widetilde{D}_m(n)$ and $D_{\rm max}(n)$ with $n$.The fewer values of $D_m$'s in  Fig.~\ref{fig:Fig5}(c) is a reflection of the  sparse  nature of the corresponding transition matrix.
The observed scaling has two implications: one, that the systems of corners for different $n$ are self-similar, and two, that one may obtain the stationary behavior for large $n$ by studying a smaller-$n$ system. This latter fact has far reaching consequences, since simulating the dynamics of squares for large $n$ typically requires use of prohibitive computational resources. Since the stationary distribution is obtained from an analysis of the transition matrix $P$, one may wonder if any signature of the aforementioned self-similarity is contained in $P$ itself. We show in Fig.~\ref{fig:Fig5}(b) and (d) that it is indeed so: the eigenvalue spectrum of $P$ for different $n$ has a `bulk' contribution that is invariant with respect to $n$ and a `tail' contribution that shifts to the right with increasing $n$. The bulk contribution contains the eigenvalue unity, and as discussed above, it is this eigenvalue that determines the stationary state. Consequently, the fact that the stationary-state distribution for different $n$ shows scaling behavior is linked to the bulk eigenspectrum being invariant with respect to $n$. The eigenvalues closer to unity are the ones that dictate the long-time behavior and hence relaxation to the stationary state~\cite{vankampen}.  For the parameter regime of  Fig.~\ref{fig:Fig5}, panels (c) and (d), the average  angular velocity $\langle \omega \rangle$ is very small. In this parameter range of  $F$  and $\mathcal{D}$,  the eigenvalue spectrum  close to unity is almost flat, i.e., the corresponding relaxation time is very large. 

\subsection{Signatures of the noise induced transition in the dynamics on the corner space}
\begin{figure}[t]
 \centering
 \includegraphics[width=.9\linewidth]{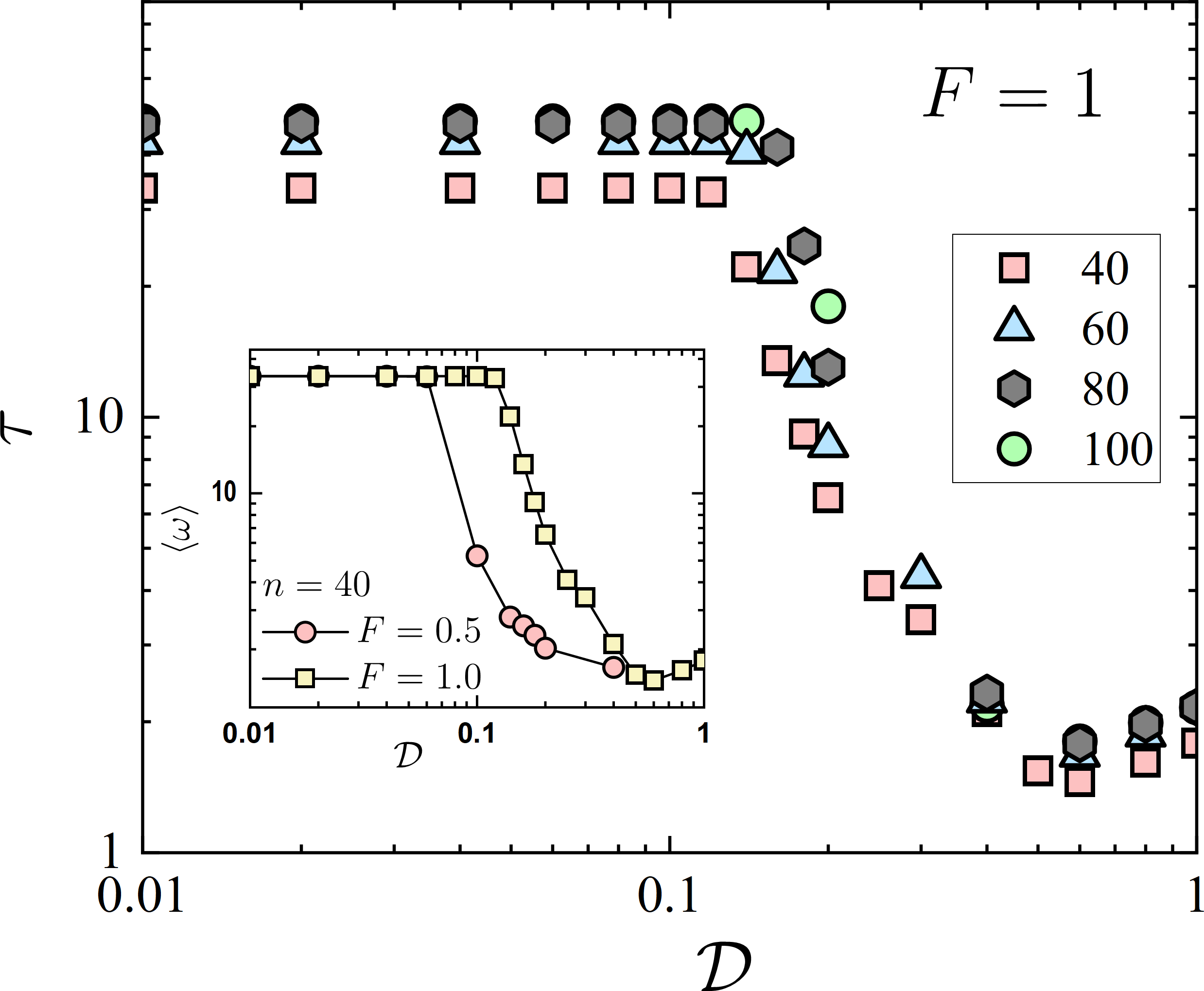}
 \caption{ The figure shows  variation of the relaxation time $\tau$ of the Markovian dynamics on the of the contacts  on the hypersurface $\{D_{0},D_{1} \ldots D_{n-1}\}$ as a function of ${\cal D}$The inset shows the  the variation $\tau$  with the noise strength $\mathcal{D}$ for  representative values of the  drive strength $F$ for $n=40$. The data for the corresponding dynamics in the configuration space  $\{\phi_1, \phi_2, \ldots \phi_n\}$ is  described in Fig. \ref{fig:current_drive}(b).}
    \label{fig:noise}
\end{figure}

The  difference of its two largest eigenvalues of the transition matrix $T$   defines the  spectral gap spectral gap $ 1-\lambda_2$ of the Markov process. The inverse of this sets the relaxation time in the system $\tau= 1/(1-\lambda_2)$.
The relaxation  time $\tau$ of Markov dynamics  constructed for the   system  described in Fig. \ref{fig:current_drive}(b) is plotted in 
Fig.~\ref{fig:noise} (b). We find  that the relaxation time $\tau$ precipitously drops as the noise strength $\mathcal{D}$ crosses a critical  value. 
This decrease in $\tau$ occurs concurrently with an increase in angular velocity $\langle \omega \rangle$ of the polygons (see Fig.~\ref{fig:current_drive}(b)).  Moreover, similar to the result for the  dynamics  in the cofiguration space (see inset of Fig.\ref{fig:current_drive} (b)), the critical noise strength at which the relaxation time $\tau$ drops   decreases to lower values with  lowering of the drive strength $F$. This is shown in the inset  to Fig.\ref{fig:noise}. Indeed,  our   description of the polygon dynamics in terms of transitions between the corners on the  hypersurface  identifies the onset of the noise induced  transition well. This reinforces our claim that the 1D coarse-grained description and the associated Markovian dynamics of the contacts  capture quite remarkably the  stationary-state behavior of the polygon dynamics which plays out in higher dimensions. 

\section{Conclusion}

In this work, we studied a one-dimensional array of curved squares undergoing noisy rotational motion about their fixed centres, and demonstrated how a coarse-grained description captures effectively the stationary-state behavior of the underlying dynamics.   Our observations reported in the paper would in principle hold true for an array of squares and in fact for any polygons.  However, in certain configurations, two  neighbouring polygons can make extended contacts where the edge of one polygon lies flat on the other. In contrast, the nature of contacts remains always point like for curved polygons. In our coarse-grained description, however, extended contacts are treated in the same footing as point contacts.  
     
Our coarse-grained description involves going from the configuration space of  squares to that of corners on the `hypersurface' that separates the accessible from the inaccessible regions of the configuration space. In this process, given a square configuration, we can always associate a unique corner configuration, but not the other way round. Consequently, one cannot obtain the dynamical trajectories of square configurations from those of the corner configurations. However, in the stationary state, if a given subset of square configurations has the highest probability of occurrence, so would be the case with the corresponding corner configurations. Then, the stationary state that one obtains with the corners is also the one of the parent system, namely, of the squares, and this explains the effectiveness of the coarse-grained description. 
     
Note that curvilinearity of polygons as measured by the parameter $\delta$ changes the structure of the allowed configuration space, as shown in Fig.~\ref{fig:Fig1n}, panels (c) -- (f). So long as $\delta$ does not break up the configuration space into isolated patches of accessible regions, our results will qualitatively be the same for different $\delta$.
     
As future directions, it would be interesting to considering asymmetric shapes or random variation between shapes generated by having a $\delta$ that is a quenched-disordered random variable along the array. Investigations in these directions are underway, and will be reported elsewhere. Our approach  maps  a  higher dimensional problem to a tractable  problem in one dimension, while retaining its essential  dynamical features.   It is worthwhile to explore the validity of this approach in  more general settings, e.g., where objects can move freely in $\mathbb{R}^3$ space, and also in situations where unlike our approach the details of the shape are not glossed over.

The authors thank  Mustansir Barma, J. Radhakrishnan and Mrinal Sarkar for useful discussions.  Shamik Gupta acknowledges support from the Science and Engineering Research Board (SERB), India under SERB-TARE scheme Grant No. TAR/2018/000023, SERB-MATRICS scheme Grant No. MTR/2019/000560, and SERB-CRG scheme Grant No. CRG/2020/000596; he also thanks ICTP – The Abdus Salam International Centre for Theoretical Physics, Trieste, Italy for support under its Regular Associateship scheme. We thank the two anonymous referees for their valuable inputs. 
	 

%

\end{document}